%% file: 00_varga_etal_SISY_v1.tex
\documentclass[conference]{IEEEtran}

\IEEEoverridecommandlockouts
\usepackage{cite}
\usepackage{amsmath,amssymb,amsfonts}
\usepackage{algorithmic}
\usepackage{graphicx}
\usepackage{textcomp}
\usepackage{xcolor}
\def\BibTeX{{\rm B\kern-.05em{\sc i\kern-.025em b}\kern-.08em
		T\kern-.1667em\lower.7ex\hbox{E}\kern-.125emX}}

\usepackage{graphicx}
\usepackage{amsmath} 
\usepackage{amssymb}  

\usepackage{color}
\usepackage{pdfpages}
\usepackage{pgfplots}
\pgfplotsset{compat=newest}
\usepackage{tikz}
\usepackage{tikzscale}
\usepackage{placeins}

\usetikzlibrary{external}

\usepackage{array}
\usepackage{booktabs}
\usepackage{multirow}
\usepackage{csquotes}
\usepackage{dblfloatfix}
\usepackage{float}
\usepackage{nicefrac}

\usepackage{enumitem}

\newcommand{\sv}[1]{\boldsymbol{#1}}

\usepackage{fancyhdr}
\fancypagestyle{firstpage}{
	\fancyhf{}

	\fancyhead[C]{\vspace{1mm} {	This work has been submitted to the IEEE for possible publication. \\ 
					Copyright may be transferred without notice, after which this version may no longer be accessible.}}
	
}

\makeatletter
\newcommand{\linebreakand}{%
\end{@IEEEauthorhalign}
\hfill\mbox{}\par
\mbox{}\hfill\begin{@IEEEauthorhalign}
}
\makeatother

\begin{document}
	\title{Cooperative Decision-Making in Shared Spaces: Making Urban Traffic Safer through Human-Machine Cooperation\\
		\thanks{This work is partly supported by the Federal Ministry for Economic Affairs and Climate Action, in the New Vehicle and System Technologies research initiative with Project number 19A21008D.}
	}
	
	\author{\IEEEauthorblockN{Balint Varga, Dongxu Yang}
		\IEEEauthorblockA{\textit{Institute of Control Systems} \\
			\textit{Karlsruhe Institute of Technology}\\
			Karlsruhe, Germany \\
			balint.varga2@kit.edu}
		\and
		\IEEEauthorblockN{Manuel Martin}
		\IEEEauthorblockA{\textit{Department Interactive Analysis and Diagnosis} \\
			\textit{Fraunhofer IOSB}\\
			Karlsruhe, Germany}
		\and
		\IEEEauthorblockN{S\"oren Hohmann}
		\IEEEauthorblockA{\textit{Institute of Control Systems} \\
			\textit{Karlsruhe Institute of Technology}\\
			Karlsruhe, Germany}
		
	}
	
	\pagenumbering{gobble} 
	\maketitle
	\thispagestyle{firstpage}
	\pagestyle{empty}
	
	\begin{abstract}
		In this paper, a cooperative decision-making is presented, which is suitable for intention-aware automated vehicle functions. With an increasing number of highly automated and autonomous vehicles on public roads, trust is a very important issue regarding their acceptance in our society. The most challenging scenarios arise at low driving speeds of these highly automated and autonomous vehicles, where interactions with vulnerable road users likely occur. Such interactions must be addressed by the automation of the vehicle. The novelties of this paper are the adaptation of a general cooperative and shared control framework to this novel use case and the application of an explicit prediction model of the pedestrian. An extensive comparison with state-of-the-art algorithms is provided in a simplified test environment. The results show the superiority of the proposed model-based algorithm compared to state-of-the-art solutions and its suitability for real-world applications due to its real-time capability.
	\end{abstract}
	
	
	\vspace*{2mm}
	\section{Introduction}
\begin{figure}[!t]
		\centering
		\includegraphics[width=0.99\linewidth]{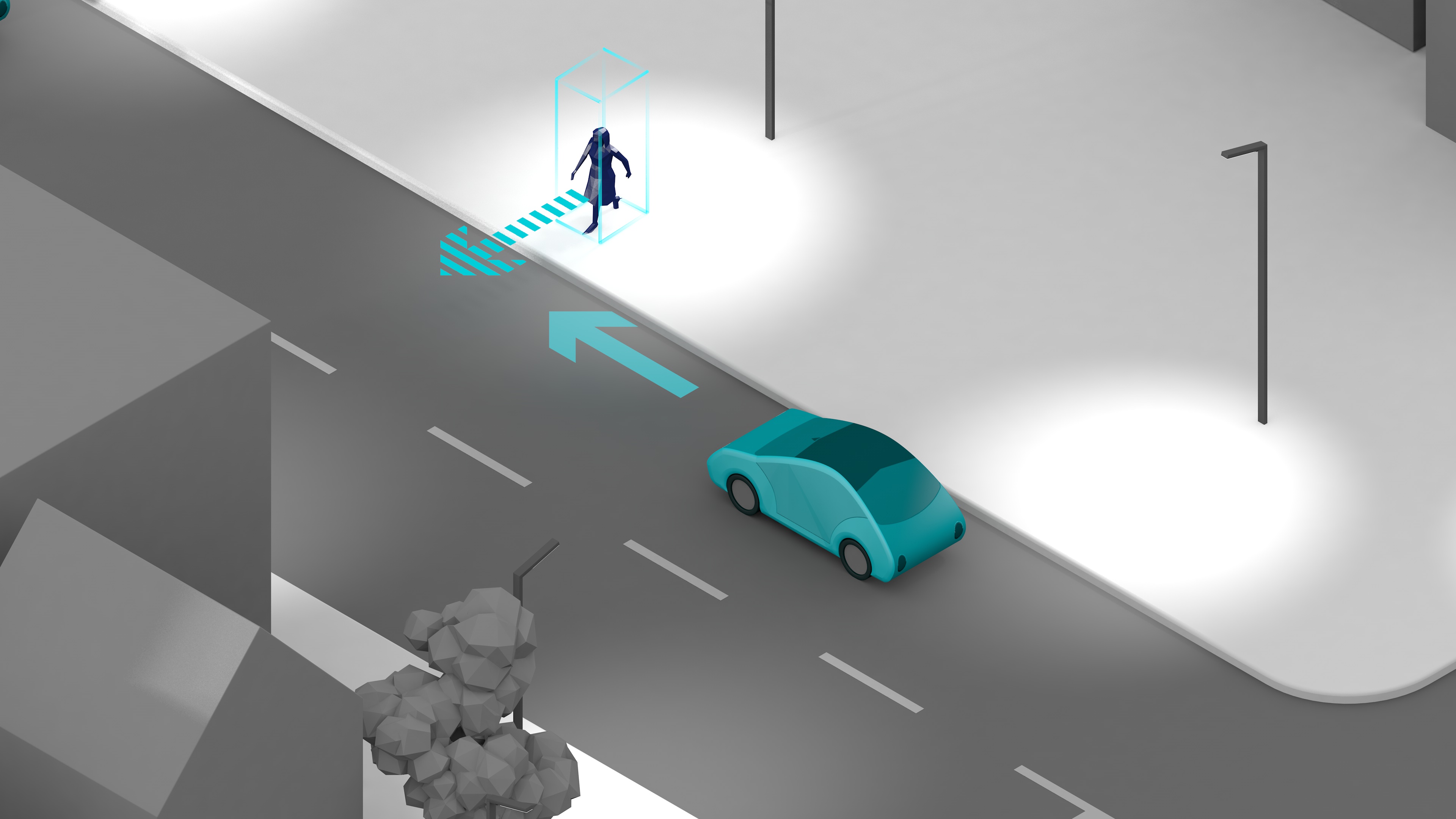}
		\caption{An example scenario, in which an interaction between the pedestrian and the automated vehicle happens. With courtesy of version1 GmbH.}
		\label{fig:example_scenario}
	\end{figure}

	Highly automated and autonomous vehicles are becoming part of our everyday life. Our society's acceptance of these systems depends on the trust of vulnerable road users (e.g. cyclists and pedestrians). Their safety is the most crucial issue~\cite{2019_PedestrianTrustAutomated_jayaraman}. For instance, accidents with automated driving functions catch the attention of the public and increase the distrust towards these systems \cite{2021_PublicAcceptancePerception_othman}. Therefore, there is extensive research to equip automated vehicles with appropriate communication channels and decision algorithms, which can handle challenging situations, e.g. urban scenarios in which vehicles have low traveling speeds and pedestrians cross the street unexpectedly~\cite{2021_VulnerableRoadUsers_tabone}. Fig.~\ref{fig:example_scenario} provides an exemplary representation of this scenario. At a low speed, an interaction between pedestrians and automated vehicles (human-machine interaction) arises, which has to be handled accordingly, leading to an increased trust in automated vehicles \cite{2021_VulnerableRoadUsers_tabone}. 
	Therefore, this paper provides the adaptation of an earlier cooperative and shared control framework \cite{2016_SharedControlSharp_flemisch, 2016_LayersSharedCooperative_pacaux-lemoine, 2019_JoiningBluntPointy_flemisch} for the problem of human-machine interaction in mixed traffic scenarios\footnote{In this paper, a mixed traffic scenario is the situation in which both a vehicle and a pedestrian can be located in the same area. Thus, the space can be shared between humans and automation.}: cooperative decision-making in shared spaces. The second contribution of this paper is the application of a novel prediction model of the pedestrian behavior, which enables explicit handling of the prediction of the pedestrian future system states and the interactions with an automated vehicle.

	The structure of this paper is the following: In Section~\ref{sec:SOA}, related works and concepts from the literature are presented and discussed. Section \ref{sec:SC-Concept} provides an adapted model-based cooperative decision-making concept with the explicit pedestrian prediction model. The simulation comparisons with former control concept are given in Section~\ref{sec:Sim_Anal}. Finally, Section~\ref{sec:Summary} provides a summary and outlook of the paper.

	\section{State of the Art} \label{sec:SOA}
	This section presents related work on the interaction problems illustrated in Fig.~\ref{fig:example_scenario}. First, problem-specific solutions are presented: Rule-based and model-based controllers for the handling of mixed intersection problems. After that, cooperative and shared control concepts for human-machine interaction are discussed.
	
	\subsection{Rule-based Controllers}
	In \cite{2015_VehiclepedestrianInteractionMixed_chao} and \cite{2019_HybridControlDesign_kapaniaa}, the vehicle-pedestrian interaction scenarios were handled by rule-based methods, 
	i.e. the decision-making happens by rules using the distance and speed of the vehicle and the human. Automated vehicle functions are adopted based on these rules to increase the trust of pedestrians. The obtained frameworks were logically verifiable since they are white-box models of the scenario. In \cite{2020_DecisionMakingProcess_hsu} and \cite{2023_IntentionAwareDecisionMakingMixed_varga}, the authors added pedestrian intention information for the controller and compared the intention-aware controller with the one without such observation. The results showed that the intention information could improve the performance of the decision-making process. Furthermore, rule-based approaches can be utilized to discover the motion and handling patterns of the pedestrian, which facilitates a better understanding of such scenarios, see~e.g.~\cite{2022_HumanMachinePatternsSystem_flemisch}. 
	
	However, the decision-making process of rule-based controllers is limited to the current time step making predictions and consideration of future situations difficult.

	\subsection{Model-based Controllers}
	Model-based controllers utilize a model for the motions of the pedestrian and the vehicle to facilitate an interaction between them. In the following, two main categories are distinguished: Model predictive controllers (MPCs) and game theoretical controllers.
	
	\vspace*{2mm}
	\subsubsection{Model Predictive Control}
	MPC has been recognized as a potentially effective method to tackle the challenges of mixed intersection scenarios. It enables real-time trajectory planning and control based on predictive models of both a vehicle and a human. In \cite{2021_AutomatedVehicleBehavior_jayaraman}, the position and speed of the vehicle and the pedestrian were used as the system states of the MPC model. The proposed algorithm was tested in a simulation environment. In \cite{2020_MultiStateSocialForce_yang, 2019_CombiningSocialForce_yang}, a novel multi-state social force based pedestrian model was proposed to predict trajectories of the pedestrian for the MPC. The objective of the MPC's took safety, efficiency, and smoothness into account. Their results showed that a smoother longitudinal velocity profile of the vehicle was generated compared to a classical proportional–integral–derivative controller. In \cite{2020_EfficientBehaviorawareControl_jayaraman}, the jerk of the vehicle was included in the cost function, which could improve the comfort during driving and has a better performance compared to a baseline rule-based controller. 
	
	However, all these proposed MPCs use implicit models of the pedestrian's motion, meaning that future interactions between the vehicle and the pedestrian are not predicted explicitly. Furthermore, the velocity of the vehicle is usually supposed to be constant, which leads to simplified decision-making algorithms.
	
	\vspace*{2mm}
	\subsubsection{Game theory}
	The interaction between two decision-making partners, like road users, can be modeled by game-theoretic mathematical tools.
	In \cite{2021_GameTheoreticApproach_amini}, a game theoretical approach for the modeling of pedestrian crossing behavior was presented, which was validated with data collected from video surveys. The model was not used by automated or autonomous vehicle functions. A negotiation model was introduced in~\cite{2020_AnalysisGameTheorybased_skugor}, which used a static game model with stochastic components. 
	A decision and a motion model were developed in \cite{2016_InteractionVehiclesPedestrians_chen}. Furthermore, an evolutionary game framework was proposed, which was calibrated and validated in measurement data from human-driven vehicles and pedestrians. In \cite{2018_WhenShouldChicken_fox}, the application of game theory for automated vehicle-human negotiation scenarios was discussed. In all the aforementioned research works, the different utility functions of the players were used to set the preferences of pedestrians and vehicles.
	
	However, these models are not suitable for the prediction of future interactions between the pedestrian and the automated vehicle, since the models used static game models
	
	\subsection{Cooperative and Shared Control Concepts}
	Cooperative and shared control concepts and frameworks are widely used to design supporting systems, see e.g. \cite{2016_SharedControlSharp_flemisch, 2019_JoiningBluntPointy_flemisch}. One of the main objectives from the cooperative and shared control literature is the haptic shared control meaning that the human and the automation interact via a haptic interface (e.g. steering wheel, joystick), see e.g. \cite{2012_SharingControlHaptics_mulder, 2018_TopologySharedControl_abbink}. These frameworks provide generic architectures, which lead to intuitive human-machine cooperation. 
	
	
	Intuitive human-machine cooperation is also crucial for mixed traffic scenarios. 
	As shown in \cite{2017_AgreeingCrossHow_rasouli}, the communication between a human driver and a pedestrian has a different nature than a haptic interaction. Thus, an adaptation of the cooperative and shared control frameworks is necessary. In some recent works \cite{2021_InfluenceRoadEnvironmental_saito}, the impact of the environment on the crossing behavior of pedestrians was addressed. However, there is no general discussion, which focuses on adapting these cooperative and shared control frameworks for mixed traffic scenarios.


	\subsection{Shortcomings of the Solutions from the Literature}
	The literature on mixed interaction scenarios with an automated vehicle does not provide a pedestrian model that can explicitly handle both the prediction of future pedestrian movement and human-machine interaction. Therefore, a model is needed that can characterize such interactions. This model should be able to predict and enable the shared decision-making of the pedestrian and the automation of the vehicle.
	
	In addition, an adaptation of the cooperative and shared control architecture is needed for a larger class of cooperative setups where there is no haptic interaction between humans and machines.
	
	This paper addresses these two research gaps by introducing a novel pedestrian model that can explicitly handle the coupling of prediction and cooperation. The model is used for implementing an MPC for the mixed scenario mentioned above. In addition, the paper discusses the elements of the general cooperative/shared control framework of~\cite{2019_JoiningBluntPointy_flemisch} and its possible adaptation to mixed traffic scenarios.

	
	

	\section{Cooperative Decision-Making for Shared Space Problems} \label{sec:SC-Concept}
\begin{figure}[!t]
		\centering
		\includegraphics[width=0.99\linewidth]{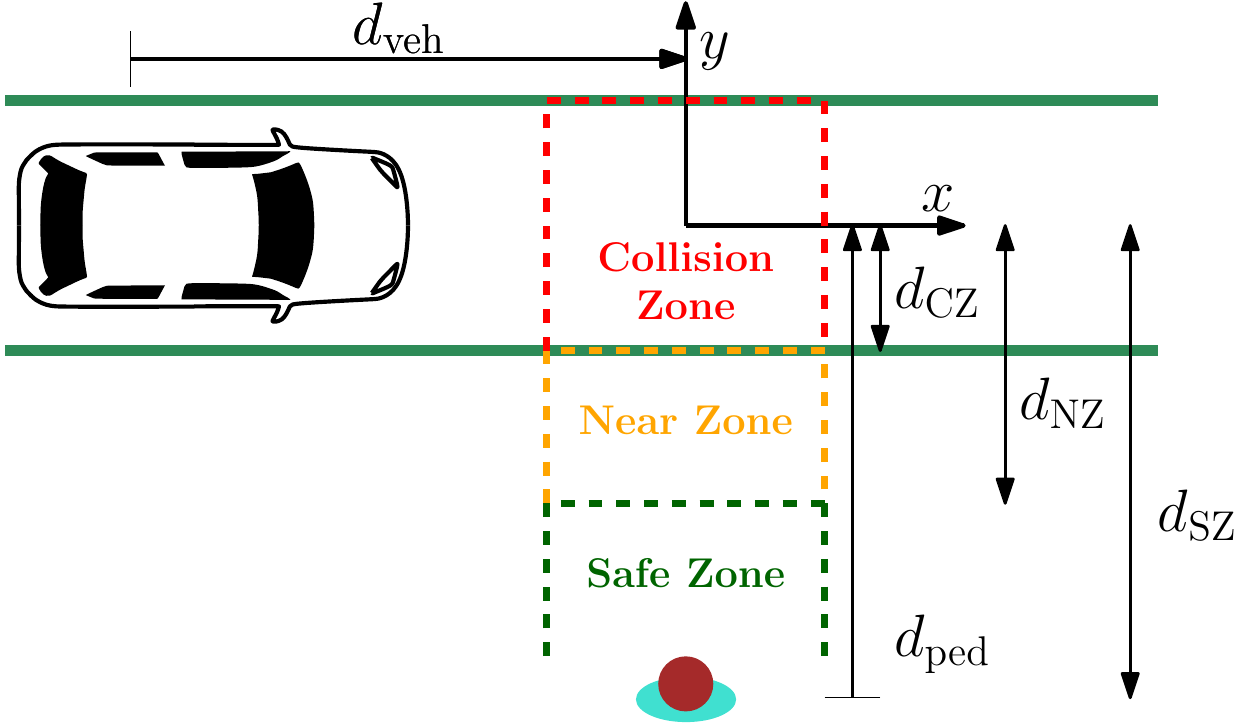}
		\caption{The top-down perspective of the automated vehicle-pedestrian interaction in a shared space}
		\label{fig:scenario_representation}
	\end{figure}

	The adapted, model-based cooperative decision-making is presented in subsequent. For the modeling of the interaction scenario, it is assumed that the pedestrian moves into the positive $y$-direction and the automated vehicle into the positive $x$-direction, only, see Fig.~\ref{fig:scenario_representation}. Note, that these assumptions do not limit the usage of the concepts, since the interaction mainly happens in these two perpendicular directions. Furthermore, the motion of the pedestrian alongside the street (parallel to the automated vehicle) is taken into account by the automated vehicle position and velocity. 
	
	\subsection{Explicit Pedestrian Model Predictive Controller}
	For the explicit motion model, it is assumed that the future velocity dynamics of the pedestrian can be modeled by 
	\begin{equation} \label{eq:human_pred_eq}
		\dot{y}_\mathrm{ped}(i+1) = \frac{1}{1 + e^{(-TTC(i)+c)}} \cdot \dot y_\mathrm{ped}^\mathrm{ref},
	\end{equation}
	where $c$ is a correction factor and depends on the pedestrian's character and the time-to-collision is computed by
	\begin{equation}
		TTC(i) = \frac{x_\mathrm{ped}(i) - x_\mathrm{veh}(i)}{\dot x_\mathrm{veh}(i)} - 
		\frac{y_\mathrm{veh}(i) - y_\mathrm{ped}(i)}{\dot y_\mathrm{ped}^\mathrm{ref}}
	\end{equation}
	and $\dot y_\mathrm{ped}^\mathrm{ref}$ is the reference velocity of the pedestrian. Note, that, in our framework, $\dot y_\mathrm{ped}^\mathrm{ref}$ is not necessarily constant. The model is explained by the following: The output of \eqref{eq:human_pred_eq} is a sigmoid function and ranges between 0 and 1. Therefore, it could be interpreted as the probability of the crossing of the pedestrian. The larger the TTC value is, the more possible is that the pedestrian would choose to walk with a reference velocity.
	
	Assuming a linear dynamics of the vehicle\footnote{This assumption is commonly utilized by the application of autonomous vehicles, see e.g. \cite{2016_RecheneffizienteTrajektorienoptimierungFur_gutjahr} or \cite[Chapter 13]{2017_ModernRoboticsMechanics_lynch}.}, the following discrete dynamic system is obtained
	\begin{align}\label{eq:explicit_dyn_model_ofMPC} \nonumber
		&\begin{bmatrix}
			x_\mathrm{veh}(i+1) \\
			\dot x_\mathrm{veh}(i+1)\\
			y_\mathrm{ped}(i+1) \\
			\dot{y}_\mathrm{ped}(i+1)
		\end{bmatrix} \\
		&\hspace*{3mm}=
		\begin{bmatrix}
			x_\mathrm{veh}(i) + \dot x_\mathrm{veh}(i) \cdot {\Delta}t + 0.5 \cdot u(i) \cdot {\Delta}t^{2} \\
			\dot x_\mathrm{veh}(i) + \Delta t \cdot u(i) \\
			y_\mathrm{ped}(i) + \Delta t \cdot \dot y_\mathrm{ped}(i) \\
			\frac{1}{1 + e^{-TTC(i)+c}} \cdot \dot y_\mathrm{ped}^\mathrm{ref}
		\end{bmatrix}_{\Huge ,}
	\end{align}
	where the desired acceleration of the vehicle $u(i)=a_\mathrm{des}$ is the system input. In order to formulate an MPC, a cost function is defined
	\begin{equation} \label{eq:MPC_explicit_cost}
		J_\mathrm{MPC} = J_\mathrm{acc} + J_\mathrm{speed} + J_\mathrm{dis},
	\end{equation}
	where
	\begin{subequations} \label{eq:J_of_MPC}
		\begin{align}
			J_\mathrm{acc} &= \sum_{i=0}^{n} w_{1} \cdot u^{2}(i), \\
			J_\mathrm{speed} &= \sum_{i=1}^{n+1} w_{2} \cdot (\dot x_\mathrm{veh}(i) - \dot x_\mathrm{veh}^\mathrm{ref}),^2 \\
			J_\mathrm{dis} &= \sum_{i=1}^{n+1} w_{3} \cdot \left(\frac{1} {\Delta x^2(i) +  \Delta y^2(i)}\right)^2
		\end{align}
	\end{subequations}
	and 
	\begin{equation} \label{eq:dist_computation}	
		\Delta x(i) = x_\mathrm{veh}(i) - x_\mathrm{ped}(i), \; \Delta y(i) = y_\mathrm{veh}(i) - y_\mathrm{ped}(i)
	\end{equation}
	hold. Using the prediction model \eqref{eq:human_pred_eq} and the cost function \eqref{eq:MPC_explicit_cost}, the following dynamic optimization is formulated to compute the inputs of the vehicle,
	\begin{subequations} \label{eq:MPC_optimization}
		\begin{align}
			\sv{u} &= \mathrm{arg}\,\mathrm{min}\, J_\mathrm{MPC} \\
			\mathrm{s.t.}\; &\text{\eqref{eq:human_pred_eq} and \eqref{eq:MPC_explicit_cost}} \\
			&d_\mathrm{min} \leq \sqrt{\Delta x^2(i) + \Delta y^2(i)} \\
			&\dot x^\mathrm{min}_\mathrm{veh} \leq \dot x_\mathrm{veh}(i) \leq \dot x^\mathrm{max}_\mathrm{veh}\\
			&a_\mathrm{veh}^\mathrm{min} \leq u(i) \leq a_\mathrm{veh}^\mathrm{max},	
		\end{align}
	\end{subequations}
	where the future inputs of the automated vehicle are aggregated into {${\sv{u} = \left[u(0), u(2),...,u(n)\right]}$}. The parameter descriptions are given in Table~\ref{table:parameters_MPC}.

	\begin{table}[!t] 
		\normalsize 
		\caption{Parameters of the model predictive shared control}
		\centering
		\begin{tabular}{c|c}
			\hline
			Symbol & Meaning\\
			\hline
			\hline
			$d_\mathrm{min}$ & Minimal safe distance \\ \hline
			$\dot{x}^\mathrm{min}_\mathrm{veh}$& Minimal necessary speed of the vehicle\\
			\hline
			$\dot{x}^\mathrm{min}_\mathrm{veh}$& Maximum allowed speed of the vehicle\\ \hline
			$a^\mathrm{min}_\mathrm{veh}$& Minimal acceleration of the vehicle\\ \hline
			$a^\mathrm{min}_\mathrm{veh}$& Maximum acceleration of the vehicle\\ \hline
			$\dot y_\mathrm{ped}^\mathrm{ref}$ &Reference velocity of the pedestrian \\ \hline
			$\dot x_\mathrm{veh}^\mathrm{ref}$ &Reference velocity of the vehicle \\
			\hline
		\end{tabular}
		\label{table:parameters_MPC}
	\end{table}

	\subsection{Intention Lowering for Deadlock Prevention}
	Deadlock, in which neither the pedestrian nor the automated vehicle move, can arise, if there is a misunderstanding between them. Fig.~\ref{fig:scenario_representation} shows the scenario from a top-down perspective. It is a usual behavior of the pedestrian that they move fast in the \textit{safe zone} and slow down near the street (in the \textit{near zone}). Sometimes, they wait in the near zone, even though they could cross. In this case, an automated vehicle would stop and wait for the pedestrian\footnote{Since safety is one of the most important aspects. Thus, a conservative behavior of automated vehicles can be expected, leading to more often stopping.}. Neither the pedestrian nor the automated vehicle move, and the interaction ends in a deadlock. In order to avoid such deadlocks, intention lowering is proposed. The intention of the pedestrian $\left(I_\mathrm{ped}\right)$ is estimated from their gestures, body posture, or eye contact by a camera system. This intention is used to modify $J_\mathrm{dis}$ in \eqref{eq:J_of_MPC} such
	\begin{equation}
		w_{3} =
		\begin{cases}
			w_{3} \cdot I_\mathrm{ped} & \text{if pedestrian is in NZ} \\
			w_{3} & \text{else}
		\end{cases}
	\end{equation}
	and the safe distance in $d_\mathrm{min}$ in (\ref{eq:MPC_optimization}c)
	\begin{equation}
		d_\mathrm{min} = 
		\begin{cases}
			d_\mathrm{min} \cdot I_\mathrm{ped} & \text{if pedestrian is in NZ} \\
			d_\mathrm{min} & \text{else}.
		\end{cases}
	\end{equation}
	
	Consequently, the automated vehicle will carefully cross the intersection after a certain time, even if the pedestrian is waiting in the near zone.

	\subsection{Adaptation of the Cooperative Control Framework}
	For our work, we adapted the cooperative control framework from \cite{2016_SharedControlSharp_flemisch} for the case of automated vehicle-pedestrian interaction in a mixed traffic scenario. Fig.~\ref{fig:flemisch} shows the framework from \cite{2016_SharedControlSharp_flemisch}, which can be used for the handling of mixed traffic scenarios. Note that in Fig.~\ref{fig:flemisch}, there is interactions at the cooperation, strategical and tactical level. On the other hand, in \cite{2016_SharedControlSharp_flemisch}, the operational level includes a haptic interaction of humans and machines.
	
	In case of an interaction between a pedestrian and an automated vehicle, the interaction has the following steps:
	\begin{figure}[!t]
		\centering
		\includegraphics[width=0.99\linewidth]{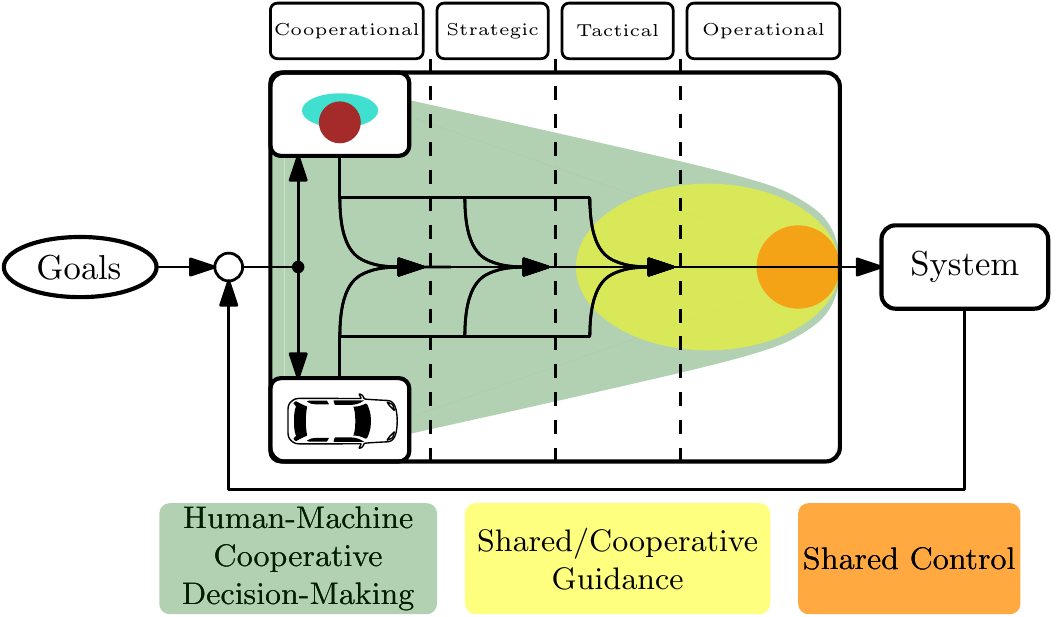}
		\caption{Shared control framework of pedestrian-automated vehicle interaction, adapted from \textit{Flemisch et. al} \cite{2019_JoiningBluntPointy_flemisch}}
		\label{fig:flemisch}
	\end{figure}

	\begin{itemize}
		\item The interaction at the \textit{cooperational} level (cf.~Fig.~\ref{fig:flemisch}) already starts in the safe zone (cf.~Fig.~\ref{fig:scenario_representation}), which is far from the intersection. Here, both the pedestrian and the automated vehicle signalize the willingness for a cooperation
		\item In the {near zone}, decisions of the \textit{strategic} and \textit{tactical} levels take place. The pedestrian and the automated vehicle cooperatively determine their future trajectories.
		\item Finally, at the \textit{operational} level, the pedestrian and the automated vehicle carry out their chosen trajectories. At this level there is no time for cooperation, merely the safety of the pedestrian can be additionally considered.
	\end{itemize}

	\section{Simulation Comparison with state-of-the-art Solutions} \label{sec:Sim_Anal}
	
	For a comparison with state-of-the-art solutions, an implicit MPC is implemented. Then, a comprehensive simulation analysis is provided in which the proposed explicit MPC is compared with the implicit MPC and the rule-based controller.

	\subsection{Model Predictive Controller with the State-of-the-Art Implicit Pedestrian}
	The baseline MPC utilizes the commonly used social force pedestrian model.
	The future states of the MPC is predicted~by
	\begin{align} \label{eq:SFM}
		\! \! \begin{bmatrix}
			y_\mathrm{ped}^\mathrm{predict}(i) \, \phantom{--} \\
			y_\mathrm{ped}^\mathrm{predict}(i+1) \\
			\vdots \\
			y_\mathrm{ped}^\mathrm{predict}(i+n)
		\end{bmatrix}
		=
		SF(x_\mathrm{veh}(i), \dot{x}_\mathrm{veh}, x_\mathrm{ped}, x_\mathrm{goal,ped}),
	\end{align}
	where $SF(\cdot)$ represents the nonlinear social force model of the pedestrian, which is widely used in literature, see e.g. \cite{1995_SocialForceModel_helbing, 2020_MultiStateSocialForce_yang, 2022_SocialInteractionsAutonomous_wang}. Consequently, the dynamic model \eqref{eq:explicit_dyn_model_ofMPC} is simplified~to
	\begin{equation}
		\small
		\begin{bmatrix}
			x_\mathrm{veh}(i+1) \\
			\dot x_\mathrm{veh}(i+1)\\
		\end{bmatrix}
		=
		\begin{bmatrix}
			x_\mathrm{veh}(i) + \dot x_\mathrm{veh}(i) \cdot {\Delta}t + 0.5 \cdot u(i) \cdot {\Delta}t^{2} \\
			\dot x_\mathrm{veh}(i) + \Delta t \cdot u(i) \\
		\end{bmatrix}
	\end{equation}
	that is used for the optimization of the MPC. Furthermore, in~\eqref{eq:dist_computation}, the distances are computed with the prediction using the positions and velocities of the pedestrian from \eqref{eq:SFM} such that
	\begin{align*} \label{eq:dist_computation_modi}	
		\Delta x(i) &= x_\mathrm{veh}(i) - x^\mathrm{predict}_\mathrm{ped}(i) \\ \Delta y(i) &= y_\mathrm{veh}(i) - y^\mathrm{predict}_\mathrm{ped}(i)
	\end{align*}
	leading to a modified value of $J_\mathrm{dis}$ in \eqref{eq:J_of_MPC}.

	\subsection{Setup of the Analysis}
	To evaluate the proposed controller, a series of simulation experiments are set up, including multiple runs, in which two controllers are compared: The proposed explicit MPC and the implicit MPC as a state-of-the-art solution. in each run, the vehicle starts from the position $(-12.5, 0)\,$m. Altogether, $100$ runs are simulated.

	For the purpose of simulation variability and to increase the fidelity of the analysis, the initial states of the vehicle and pedestrian are perturbed as follows:
	\begin{align*}
		x_{\mathrm{ped}}^{\mathrm{init}} &= 0.0 + \mathcal{N}(0, 1) \; \mathrm{m},\\
		y_{\mathrm{ped}}^{\mathrm{init}} &= \max(3.5 + \mathcal{N}(0, 0.5),  2.0)  \; \mathrm{m},\\
		\dot y_{\mathrm{ped}}^{\mathrm{init}} &= 1.4 + \mathcal{N}(0, 0.1)  \; \mathrm{m/s}, \\
		\dot x_{\mathrm{veh}}^{\mathrm{init}} &= 6.0 + \mathcal{N}(0, 0.5)  \; \mathrm{m/s}.
	\end{align*}
	Furthermore, the intention of the pedestrian is perturbed as well
	\begin{equation}
		I_{\mathrm{ped}} \sim \begin{cases}
			\mathcal{U}(0.5, 1.0) & \text{if pedestrian intents to cross,} \\
			\mathcal{U}(0.0, 0.5) & \mathrm{else.}
		\end{cases}
	\end{equation}
	The function $\mathcal{N}\left(a,b\right)$ represents the continuous normal distribution, and the function $\mathcal{U}\left(a,b\right)$ the continuous uniform distribution in the interval $ a,b$. The simulation model of the pedestrian in the simulation is an SFM. Since detailed vehicle characterization is not necessary for our negotiation scenario, a linear vehicle model is used.

	Once, the vehicle passes the interaction, the simulation run is finished. Furthermore, in case of collisions or violation of the time threshold, the simulation is terminated. The measure for the evaluation of the proposed cooperative decision-making algorithm is composed of 
	\begin{itemize}
		\item The minimal time to collision ($TTC_\mathrm{min}$): The smallest $TTC$ value during the simulation run.
		\item Total simulation time $t_{Tot}$: The time is necessary for the vehicle to pass the intersection.
		\item Maximal acceleration $\left|a_\mathrm{max}\right|$: The absolute maximum acceleration of the vehicle during the simulation run, which gives information about the driving comfort.
	\end{itemize}
	The overall score for each run is a weighted summation of these three factors:
	\begin{equation} \label{eq:overall_goal}
		J_\mathrm{overall}=k_{1} \cdot TTC_\mathrm{min} - k_{2} \cdot t_\mathrm{Tot} - k_{3} \cdot \left|a_\mathrm{max}\right|
	\end{equation}
	In the analysis, $k_{1}=k_{2}=k_{3}=1$ were set. Furthermore, in case of a collision between a vehicle and a human, the overall score $J_\mathrm{overall}$ is penalized additionally. The average score from the 100 runs is used for the assessment. 
	
	The tuning of the controllers is carried out by the \textit{Optuna} optimization framework \cite{2019_OptunaNextgenerationHyperparameter_akiba} and by the objective function~\eqref{eq:overall_goal}, which lead to the optimal values of $w_{1}$, $w_{2}$, $w_{3}$ in \eqref{eq:J_of_MPC}. Thus, there is no bias due to a manual tuning of the controller
	
	Two evaluations with a model of the pedestrian are conducted: In the first, an SFM is used. In the second, a mixed human motion model is used: This means that either an SFM or constant speed human motion model\footnote{This simple characterization is interpreted as the non-rational behavior of the pedestrian.} is chosen randomly. In order to handle such behavior, for the second run, the two MPCs have an additional prediction with the constant speed human motion model, which leads to an increased computation time.
	
	Finally, a manual test is conducted, in which the decision-making of the pedestrian is based on the inputs of a test person. A test person can set and adjust the velocities and the intention of the pedestrian in the simulation, using a graphical user interface and keyboard inputs. This leads to a more realistic behavior and a higher validity of the results.
	\subsection{Results}

\begin{figure}[!t]
		\centering
		\includegraphics[width=0.99\linewidth,height=4.75cm]{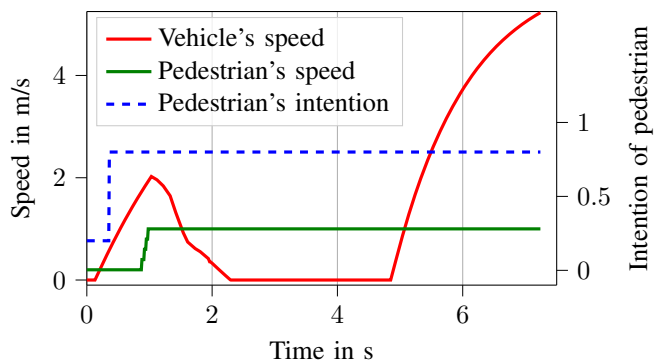}
		\caption{Simulation results of the scenario with decision of a test person is illustrated, in which the pedestrian crosses before the vehicle.}
		\label{fig:res_sim1}
	\end{figure}

	The results of the first evaluation are presented in Table~\ref{table:evaluation with simple human model and simple controller}. It can be seen that the explicit MPC outperforms the other two state-of-the-art solutions. Furthermore, the computation time of all three controllers are small enough for real-world usage. In the second evaluation, the two MPCs have poorer results compared to the first evaluation, see Table~\ref{table:evaluation with complex human model and complex controller}. The reason for that is the mixing of the human model, which leads to a more realistic setup and a slightly higher computational time due to the additional prediction.
	\begin{table}
		\centering
		\normalsize
		\caption{Evaluation with SFM of the pedestrian. \\ Note a larger value indicating a better result.}
		\begin{tabular}{c|c c c}
			\hline
			Controller & Rule-based & \begin{tabular}{@{}c@{}}Implicit\\ MPC\end{tabular} & \begin{tabular}{@{}c@{}}Explicit\\ MPC\end{tabular}\\
			\hline 
			&&&\\[-1em]
			$J_\mathrm{overall}$ & -16.50 & -4.52 & -1.22 \\ 
			&&&\\[-1em]
			\hline
			\begin{tabular}{@{}c@{}}Average  \\ Comp. Time\end{tabular} & 1.5e-6 & 0.008 & 0.009 \\
			\hline
		\end{tabular}
		\label{table:evaluation with simple human model and simple controller}
	\end{table}
	\begin{table}
		\centering
		\normalsize
		\caption{Evaluation with the mixed human motion model. \\ Note a larger value indicating a better result.}
		\begin{tabular}{c|c c c}
			\hline
			Controller & Rule-based & \begin{tabular}{@{}c@{}}Implicit\\ MPC\end{tabular} & \begin{tabular}{@{}c@{}}Explicit\\ MPC\end{tabular}\\
			\hline
			&&&\\[-1em]
			$J_\mathrm{overall}$ & -16.01 & -7.24 & -3.23 \\ 
			&&&\\[-1em]
			\hline
			\begin{tabular}{@{}c@{}}Average  \\ Comp. Time\end{tabular} &  1.5e-6 & 0.016 & 0.017 \\
			\hline
		\end{tabular}
		\label{table:evaluation with complex human model and complex controller}
	\end{table}
	Finally, the results with the test person are shown in Fig.~\ref{fig:res_sim1}, in which the pedestrian crosses the intersection before the automated vehicle. The velocity and the intention of the pedestrian are generated by a test person, which can not be described by an SFM or other pedestrian models. It can be observed: At the beginning, the intention of the pedestrian does not correspond with their velocity and the vehicle does not decelerate. Only after, the velocity of the pedestrian is increased, the vehicle decelerates and let the pedestrian cross the street.
	
	The analysis presented in this paper provides promising results for the real-world application of the proposed explicit MPC. Since no collision has arisen in the analysis and the computational time is sufficiently small, real-world tests will be possible. However, the performance of MPC depends strongly on the accuracy of the prediction of human trajectory. Therefore, in our further work, the human motion analysis needs to be addressed and to be verified whether the model of the MPC controller provides a sufficient prediction.  
	
	\vspace*{-0.75mm}
	\section{Summary and Outlook} \label{sec:Summary}
	\vspace*{-0.5mm}
	This paper proposed a cooperative decision-making of an automated vehicle for mixed traffic scenarios, in which a pedestrian intents to cross the street through an unsignalized intersection. The automated vehicle takes into account the intention of the pedestrian and negotiates the situation of whether to stop or not before the pedestrian's crossing. The novelties of the paper are the explicit pedestrian model and adaptation of the cooperative and shared control framework for mixed traffic scenarios. The results showed that the novel model outperforms the state-of-the-art rule-based controller and MPC using a social force model. 
	In our future work, we plan to test cooperative decision-making using real automated vehicles and high-fidelity human-in-the-loop studies.

	\vspace*{-0.75mm}
	\bibliographystyle{IEEEtran}
	\input{00_varga_etal_SISY_v1.bbl}


\end{document}